\documentclass[aps,pra,showpacs,twocolumn,amsmath,amssymb,superscriptaddress,floatsfix,nofootinbib]{revtex4}
\usepackage{times}
\usepackage[dvips]{graphicx}
\usepackage{color}
\usepackage{booktabs}
\usepackage{dcolumn,bm}
\renewcommand{\d}{\mathrm{d}}

\newcommand{\be}{\begin{equation}}
\newcommand{\ee}{\end{equation}}
\newcommand{\bea}{\begin{eqnarray}}
\newcommand{\eea}{\end{eqnarray}}
\newcommand{\bse}{\begin{subequations}}
\newcommand{\ese}{\end{subequations}}

\newcommand{\kf}{k_{\mathrm F}}
\newcommand{\ef}{\varepsilon_{\mathrm F}}
\newcommand{\mathsym}[1]{{}}

\begin{document}

\title{Decay of superfluid currents in the interacting 
one-dimensional Bose gas}

\author{Alexander Yu.~Cherny}
\affiliation{Bogoliubov Laboratory of Theoretical Physics, Joint Institute for Nuclear
Research, 141980, Dubna, Moscow Region, Russia}

\author{Jean-S\'ebastien Caux} 
\affiliation{Institute for Theoretical Physics, University of Amsterdam, 1018 XE 
Amsterdam, The Netherlands}

\author{Joachim Brand}
\affiliation{Centre for Theoretical Chemistry and Physics and Institute of Natural
Sciences, Massey University, Private Bag 102 904, North Shore, Auckland 0745, New Zealand}

\date{July 28, 2009}

\begin{abstract}
We examine the superfluid properties of a one-dimensional (1D) Bose gas in a ring trap based on the model 
of Lieb and Liniger. While the 1D Bose gas has nonclassical rotational inertia and 
exhibits quantization of velocities, the metastability of currents depends sensitively on 
the strength of interactions in the gas: the stronger the interactions, the faster the 
current decays.  It is shown that the Landau critical velocity is zero in the 
thermodynamic limit due to the first supercurrent state, which has zero energy and 
finite probability of excitation. We calculate the energy dissipation rate of ring 
currents in the presence of weak defects, which should be observable on experimental 
time scales. 
\end{abstract}


\maketitle

\section{Introduction}

Superfluidity is one of the most dramatic manifestations of quantum mechanics on the 
macroscopic scale, and is associated to a host of different phenomena such as 
non-classical rotational inertia, quantization of vortices, dragless motion of 
impurities and metastability of ring currents as seen in, {\it e.g.}, liquid He II. Since 
each of these phenomena may be taken as ``defining'' a transition to superfluidity, it 
is important to ask under what circumstances they occur together. As was pointed out by 
Leggett \cite{leggett99} the metastability of ring currents and nonclassical rotational 
inertia are two fundamental superfluid phenomena of yet very different nature. While the 
latter is an equilibrium property, the former is a dynamic one. Although both types of 
phenomena are often explained by Bose-Einstein condensation of bosons or Cooper 
pairs of fermions \cite{leggett99}, the latter is not seen as an exclusive requirement 
\cite{sonin70,leggett73}. Here we consider the superfluid properties of an interacting 
one-dimensional (1D) Bose gas at zero temperature, a system which is not Bose-condensed 
\cite{bogoliubov61:book,hohenberg67} but may possess quasi-long-range order 
\cite{petrov00}. It is a long-standing question whether the 1D Bose gas can support 
persistent currents with macroscopic lifetimes \cite{hohenberg67}.

This system has been realized with ultracold bosonic atoms in tightly confining linear 
traps \cite{Kinoshita2004,palzer09} (ring traps are also under development 
\cite{gupta05:ring_BEC}), in which the boson interactions are effectively described 
\cite{olshanii98,cherny04} by the contact potential $V(x)=g_{\mathrm B}\delta(x)$ of the 
Lieb-Liniger (LL) model \cite{lieb63:1}. The interaction strength is quantified by the 
dimensionless parameter $\gamma= m g_{\mathrm{B}}/(\hbar^2 n)$, where $n$ is the linear 
density and $m$ is the mass. For $\gamma \rightarrow \infty$, the model is known as the 
Tonks-Girardeau (TG) gas and can be mapped to an ideal \emph{Fermi} gas. For $\gamma 
\ll 1$, the Bogoliubov model of weakly interacting bosons is recovered.

Experimental investigation of the superfluid properties of the 1D Bose gas by observing 
the motion of impurities is at an early stage \cite{palzer09} and theoretical 
predictions are not yet comprehensive.  Sonin \cite{sonin70} found that ring currents 
can be metastable except for infinitely strong interactions. Kagan {\it et al.} 
\cite{kagan00} also concluded  that persistent currents could be observable on 
experimental time scales and B\"uchler {\it et al.} \cite{buchler03} found the 1D Bose 
gas able to sustain supercurrents even in the presence of a strong defect. Astrakharchik 
and Pitaevskii \cite{astrakharchik04} considered the drag force on a moving heavy 
impurity within Luttinger liquid theory and predicted a power-law dependence on the 
velocity for small velocities. These results contain an unknown prefactor preventing the 
calculation of the actual value of the drag force and are in any case not applicable at 
larger velocities. The motion of an impurity of finite mass was considered in the TG gas 
\cite{girargeau09} but for finite values of $\gamma$ this problem is still unresolved.

\begin{figure}[tbh]
\centerline{\includegraphics[width=1.\columnwidth]{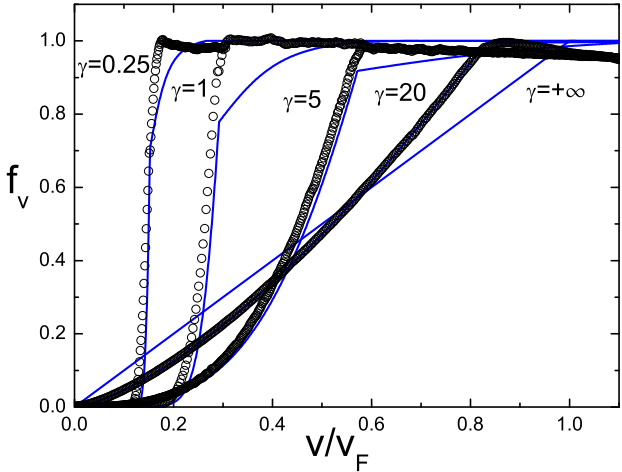}} 
\caption{\label{fig:drag} (Color online) The dimensionless drag force versus the 
velocity (relative to $v_{\rm F} = \hbar \pi n/m$) of the impurity at various values of 
the coupling parameter. The solid (blue) lines represent the force obtained with 
Eqs.~(\ref{dragf1D}) and (\ref{sq_our}); open circles are the numerical data obtained 
using ABACUS \cite{caux06}.}
\end{figure}

In this paper we calculate the rate of energy dissipation of ring currents in the 
presence of a small integrability-breaking defect of strength $g_i$ based on recent 
advances in the understanding \cite{brand05,caux06,imambekov08} of the dynamics of the 
LL model. The results of our calculations are summarized in Fig.~\ref{fig:drag}. While 
for small velocities our calculations support the power-law predictions of 
Ref.~\cite{astrakharchik04}, the drag force $F_{\rm v} = 2g_{\rm i}^2 n m \hbar^{-2} 
f_{\rm v}$ assumes the  velocity-independent value of $2g_{\rm i}^2 n m /\hbar^2$ for 
velocities large compared to the speed of sound $c$. Although our results suggest that 
the 1D Bose gas can support metastable currents only in the weakly interacting regime 
where $\gamma \ll 1$, the superfluid fraction is 1 regardless of $\gamma$ \cite{ueda99} 
according to the nonclassical rotational inertia for a finite ring.

\section{Landau criterion of superfluidity} 

\begin{figure}[tbh]
\includegraphics[width=1.\columnwidth]{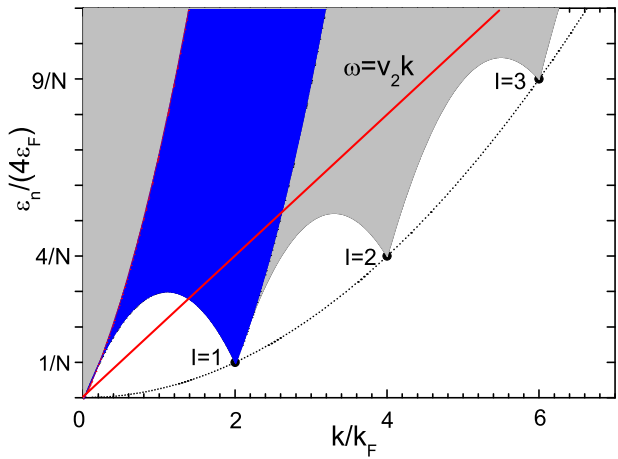}
\caption{\label{fig:schem} (Color online) Schematic of the excitation spectrum of the 1D 
Bose gas in a perfectly isotropic ring. The supercurrent states $I$ lie on the parabola 
$\hbar^2k^2/(2 N m)$ (dotted line). Excitations occur in the shaded area; the discrete 
structure of the spectrum is not shown for simplicity. The blue (dark) area represents 
particle-hole excitations \cite{lieb63:2}. Motion of the impurity with respect to the 
gas causes transitions from the ground state to the states lying on the straight (red) 
line. }
\end{figure}

In the LL model the total momentum is a good 
quantum number, and periodic boundary conditions quantize it in units of $2\pi\hbar/L$, 
where $L$ is the ring circumference. The low-lying spectrum of $N=nL$ bosons as shown in 
Fig.~\ref{fig:schem} has local minima \cite{haldane81} at the supercurrent states $I$ 
($I=0,1,2,\ldots$) with momenta $p_I = 2 \pi n \hbar I$ and excitation energies 
$\varepsilon_I = p_I^2/(2 N m)$. These correspond to Galilean transformations of the 
ground state with velocities $v_I = p_I/(N m)$. The minima do not depend on interactions 
and tend to zero in the limit of large system size.

Suppose that the gas is initially rotating with the linear velocity $-v_I$ and then is 
braked with an ``obstacle,'' created, {\it e.g.}, by a laser beam 
\cite{raman99:critical_velocity}. In the frame where the gas is at rest, the obstacle 
moves with velocity $v_I$. In a superfluid we expect no energy dissipation, and thus 
zero drag force (the current is persistent). Energy conservation dictates that the 
transitions from the ground state caused by the moving obstacle with velocity $v$, lie 
on the line $\varepsilon= v p$. According to Landau, if the excitation spectrum lies 
above this line, the motion cannot excite the system, which is then regarded superfluid. 
The Landau critical velocity (when the line touches the spectrum) equals $v_\mathrm{c}= 
\varepsilon_1/p_1=v_1/2$. This implies that any supercurrent state with $I\geqslant1$ is 
unstable since $v_I>v_\mathrm{c}$. However, in 3D similar supercurrent states exist, 
which apparently leads to the absence of current metastability. The paradox can be 
resolved by considering not only the spectrum but also \emph{probabilities} of 
excitations. Below we argue that in the 3D case, the probability to excite supercurrents 
is vanishingly small, while in the 1D case it depends on the strength of bosonic 
interactions. 

\section{Hess-Fairbank effect}

When the walls of a toroidal container are set in rotation 
adiabatically with a small velocity, a superfluid stays at rest while a normal fluid 
follows the container. This effect leads to a nonclassical rotational inertia of 
superfluid systems, which can be used to determine the superfluid fraction 
\cite{leggett73}. One can show \cite{ueda99} that in the LL model, the gas 
has zero rotational inertia (zero normal fraction) at $T=0$ for any $\gamma>0$. This is 
an {\it equilibrium} property completely determined by the low-lying energy spectrum 
\cite{leggett99}.

\section{Decay of supercurrents}

\subsection{Dynamic response} 

By contrast to the Hess-Fairbank effect, metastability of currents is not an equilibrium 
effect and transition probabilities have to be considered. The dissipation rate as 
energy loss per unit time $\dot{E}$ of an obstacle (or heavy impurity) moving with 
velocity $v$ relative to the gas can be related to the drag force $F_{\mathrm{v}}$ 
acting on the impurity by $\dot{E}=-F_{\mathrm{v}} v$. For weak impurities with 
interaction potential $V_\mathrm{i}(x)$ the drag force is related to the dynamic 
structure factor (DSF) in linear-response theory \cite{astrakharchik04,timmermans98}:
\begin{equation}
F_{\mathrm{v}}(v) =\int_{0}^{+\infty}\d k\,k |\tilde{V}_\mathrm{i}(k)|^2 S(k,k v)/L,
\label{dragf1D}
\end{equation}
where $\tilde{V}_\mathrm{i}(k)$ is the Fourier transform of the impurity potential. The 
DSF $S(k,\omega)$ describes the transition probability between the ground state 
$|0\rangle$ and excited states $|m\rangle$ with energy transfer $\hbar \omega$ and 
momentum transfer $\hbar k$ caused by a density perturbation, and can be written as 
\begin{equation}
S(k,\omega)=\sum_m |\langle0|\delta\hat{\rho}_k|m\rangle|^2\delta(\hbar\omega-E_m+E_0), 
\label{eqn:dsfenergy}
\end{equation}
where $\delta\hat{\rho}_k=\sum_{j}e^{-i k x_j}-N\Delta(k)$ is the Fourier component of 
the density operator, $\Delta(k)=1$ at $k=0$ and $\Delta(k)=0$ otherwise. Several 
results for the DSF in the LL model have recently become available 
\cite{brand05,caux06,imambekov08}. It can be measured in cold gases by  Bragg scattering 
\cite{stenger99,ozeri04}. 

\begin{figure}[h]
\centerline{\includegraphics[width=1.\columnwidth,clip=true]{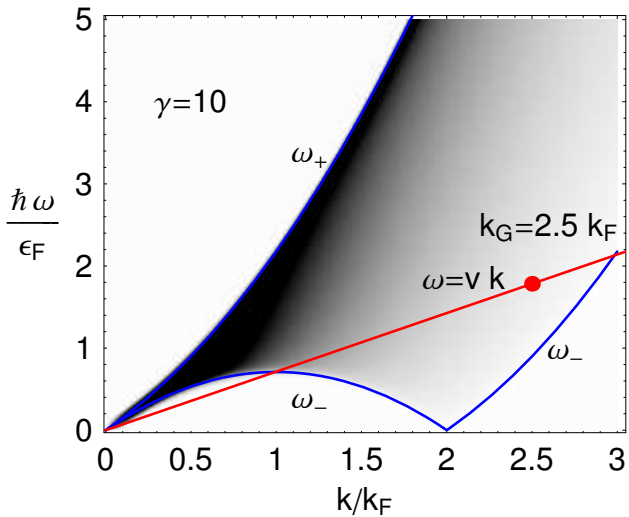}} 
\caption{\label{fig:sqom3D_01} (Color online) Dynamic structure factor of the 1D Bose 
gas from \cite{caux06} for $N=100$. Dimensionless values of $S(k,\omega)\ef/N $ are 
shown in shades of gray between 0 (white) and 0.7 (black). The full (blue) lines 
represent the limiting dispersion relations $\omega_\pm$ and the straight (red) line is 
the line of integration in Eq.~(\ref{dragf1D}). Only one point at $k=k_G$, shown in full 
(red) circle, contributes to the integral when the perturber is a shallow cosine
potential with a reciprocal vector $k_G$.}
\end{figure}
Numerical values of DSF calculated with the ABACUS algorithm \cite{caux06} are shown in 
Fig.~\ref{fig:sqom3D_01}. The probability to create multiparticle excitations lying 
outside of the region $\omega_{-}(k)\leqslant\omega\leqslant\omega_{+}(k)$ are 
identically zero (below $\omega_-$) or very small (above $\omega_+$). Transitions from 
the ground state caused by a moving obstacle with velocity $v$ occur along the straight 
(red) line. Drag force (\ref{dragf1D}) is thus a generalization of the Landau 
criterion for superfluidity. Indeed, if the excitation spectrum of a generic system lies 
above the line $\omega= v k$ then it is superfluid; in this case the drag force 
(\ref{dragf1D}) equals zero. The drag force thus proves to be fundamental and can be 
considered as a quantitative measure of superfluidity.

\subsection{Shallow optical lattices}

Equation (\ref{dragf1D}) can be verified experimentally for 
different types of obstacles:  for $V_\mathrm{i}(x)=g_\mathrm{i}\delta(x)$ all the 
points at the line contribute to the drag force, while for 
$V_\mathrm{i}(x)=g_\mathrm{L}\cos(2\pi x/a)$ only one point $(k_G,k_G v)$ in the 
$k$-$\omega$ plane does, where $k_G\equiv 2 \pi/a$ is the reciprocal lattice vector (see 
Fig.~\ref{fig:sqom3D_01}). Indeed, substituting the Fourier transform into 
Eq.~(\ref{dragf1D}) yields
\begin{equation}
F_{\mathrm{v}} =\pi g_{\mathrm{L}}^{2}k_G S(k_G,k_G v)/2.
\label{dfkG}
\end{equation}
The filling factor of the lattice is given by $2\pi n/k_G$. Equation (\ref{dfkG}) can be 
exploited even in the case of a cigar-shaped quasi-1D gas of bosons at large number of 
particles, because the boundary conditions do not play a role in the thermodynamic 
limit. It gives us the momentum transfer per unit time from a moving shallow lattice, 
which can be measured experimentally \cite{fallani04}. At $k_G=2\pi n$, corresponding to 
the Mott insulator state in a deep lattice, and at $\gamma \gg1$, the drag force takes 
non-zero values for arbitrary $v\leqslant \omega_+(k_G)/k_G$. However, at small 
$\gamma$, its non-zero values practically localize  in vicinity of 
$v=\omega_+(k_G)/k_G$. As there is no sharp transition from superfluid to isolated phase 
in 1D \cite{polkovnikov05}, we can put the threshold equal to, say, $0.1$ of the 
characteristic value $\pi g^2_\mathrm{L}k_G N/(8\,\ef)$ of the drag force (\ref{dfkG}). Then we get 
a phase diagram in the $v$-$\gamma$ plane \cite{cherny09a} similar to that of Polkovnikov 
\emph{et al.} \cite{polkovnikov05}. Note that in the latter paper, the superfluidity was 
examined in terms of quantum phase slips \cite{arutyunov08}. So, the both quasiparticle and 
quantum phase slip description lead to the same results.

\subsection{Approximate expression for drag force}

In order to study metastability of the $I$th supercurrent state, we need to calculate 
the drag force on an obstacle moving with the velocity $v_I$ relative to the gas. For 
large system size the supercurrent-state velocities are dense and in the thermodynamic 
limit ($N\to \infty$, $n =\mathrm{const}$) we may consider arbitrary velocities. We 
consider the drag force and decay of currents in various regimes.

We calculate the drag force from Eq.~(\ref{dragf1D}) by using the interpolating expression
\begin{equation}
\label{sq_our}
S(k,\omega)=C (\omega^{\alpha}-\omega_{-}^{\alpha})^{\mu_{-}}/(\omega_{+}^{\alpha}-\omega^{\alpha})^{\mu_{+}}
\end{equation}
for $\omega_{-}(k)\leqslant\omega\leqslant\omega_{+}(k)$, and $S(k,\omega)=0$ otherwise 
\cite{cherny08}. Here, $K\equiv \hbar \pi n/(m c)$ is the Haldane parameter 
\cite{astrakharchik04}, $\mu_{+}(k)$ and $\mu_{-}(k)$ are the exact exponents 
\cite{imambekov08} at the borders of the spectrum $\omega_{+}(k)$ and $\omega_{-}(k)$, 
and  $\alpha\equiv 1+1/\sqrt{K}$. The values of $\omega_{\pm}(k)$ and $\mu_{\pm}(k)$ can 
be calculated from the coupling constant $\gamma$ numerically by the methods outlined in 
Refs.~\cite{lieb63:2,imambekov08}. The normalization constant $C$ depends on momentum 
but not on frequency and is determined from the $f$-sum rule $\int_{-\infty}^{+\infty} 
\d\omega\, \omega S(q,\omega)= N{q^{2}}/{(2m)}$. The expression (\ref{sq_our}) is 
applicable for all ranges of the parameters $k$, $\omega$, and $\gamma$ with increasing 
accuracy at large $\gamma$. A more detailed discussion can be found in 
Ref.~\cite{cherny08}.

\subsection{Numerical results}

We further restrict ourselves to a $\delta$-function impurity interaction with 
$\tilde{V}_\mathrm{i}(k)=g_\mathrm{i}$. Results of integrating Eq.~(\ref{dragf1D}) are 
shown in Fig.~(\ref{fig:drag}). For large velocities  the drag force reaches the 
velocity-independent value of $2g_{\rm i}^2 n m /\hbar^2$. A characteristic velocity 
scale is the speed of sound $c$, which determines the transition from a power-law 
increase to the velocity independent regime. The speed of sound $c$ of the LL model is 
proportional to $\gamma$ for small $\gamma$ but saturates to the value $v_{\rm F}$ for 
large $\gamma$ \cite{lieb63:1}. The numerical DSF as per Ref.~\cite{caux06} was obtained 
for $N = 150$ particles ($\gamma = 20, 5$), $N = 200$ ($\gamma = 1$) and $N = 300$ 
($\gamma = 0.25$). The $f$-sum rule saturations at $k = 2k_F$ were $99.64\% (\gamma = 
20)$, $97.81\% (\gamma = 5)$, $99.06\% (\gamma = 1)$, and $99.08\% (\gamma = 0.25)$, with 
yet better results at smaller momentum.  The fit with the analytical ansatz is good for 
all values of $v$ for large $\gamma$.  The decreasing curves at large velocity $v \gg c$ 
are due to imperfect sum-rule saturation at high momenta. For small $\gamma$, the onset 
of the drag force is quicker from the numerical DSF than from the analytical ansatz.  
This occurs first because the smoothing of the numerical data required to compute the 
drag force overestimates it when its curvature is positive (this smoothing also leads to 
small artifacts in the data around $v = c$), and second because the obtained numerical 
DSF is larger than the analytical ansatz for $\omega \ll \omega_+$, and also just above 
the Bogoliubov dispersion (where the analytical ansatz is zero by definition), where 
excitations with higher numbers of particle-hole pairs contribute.

\subsection{Drag force at small velocities}

For the important question whether persistent currents may exist at all, the small 
velocity regime is most relevant, which is dominated by transitions near the first 
supercurrent state (umklapp point at $k = 2k_{\rm F}$). The drag force in this regime 
has a power-law dependence on the velocity $F_\mathrm{v}\sim v^{2K-1}$ for $v\ll c$, as 
first found by Astrakharchik and Pitaevskii. From Eqs.~(\ref{dragf1D}) and 
(\ref{sq_our}) we can obtain:
\begin{align}
&f_\mathrm{v}\equiv\frac{F_\mathrm{v}\pi\ef}{g_\mathrm{i}^2 \kf^3} \simeq 2K 
\bigg(\frac{v}{v_\mathrm{F}}\bigg)^{2K-1}\bigg(\frac{4\ef}{\hbar\omega_{+}(2\kf)}\bigg)^{2K}\nonumber \\ 
&\times\!\!\frac{\Gamma\Big(1+\dfrac{2 K}{\alpha}-\mu_{+}(2\kf)\Big)}{\Gamma\Big(\dfrac{2 K}{\alpha}\Big)\Gamma\big(1-\mu_{+}(2\kf)\big)}
 \frac{\Gamma\big(1+\mu_{-}(2\kf)\big)\Gamma\Big(1+\dfrac{1}{\alpha}\Big)}{\Gamma\Big(1+\mu_{-}(2\kf)+\dfrac{1}{\alpha}\Big)},
\label{DF_universal1}
\end{align}
where $\Gamma(x)$ is Euler's gamma-function, and $\mu_{-}(2\kf) 
=2\sqrt{K}(\sqrt{K}-1)$~\cite{imambekov08}. This formula is valid for {\it arbitrary} 
coupling constant and works even in the Bogoliubov regime at $\gamma\ll 1$. In practice, 
Eq.~(\ref{DF_universal1}) works well up to $v\lesssim 0.1 c$.

\subsection{Why excitations near the umklapp point do not play a role in three dimension}

The behavior of the DSF near the umklapp point means that the drag force takes non-zero 
values even for {\em arbitrarily small} interactions. This fact is related to the 
absence of Bose-Einstein condensation in the 1D Bose gas. For large interactions, 
umklapp excitations become readily available and provide an avenue for the rapid decay 
of supercurrents. Landau  reasoned that  this is very implausible in 3D, since it 
involves the macroscopic motion of the system, and hence a {\it macroscopic number} of 
quasiparticle excitations. Strong correlations in 1D, however, make umklapp excitations 
easily accessible, since they involve only a {\it single} fermionic-like quasiparticle 
\cite{lieb63:1}.

\begin{figure}[h]
\centerline{\includegraphics[width=1.\columnwidth]{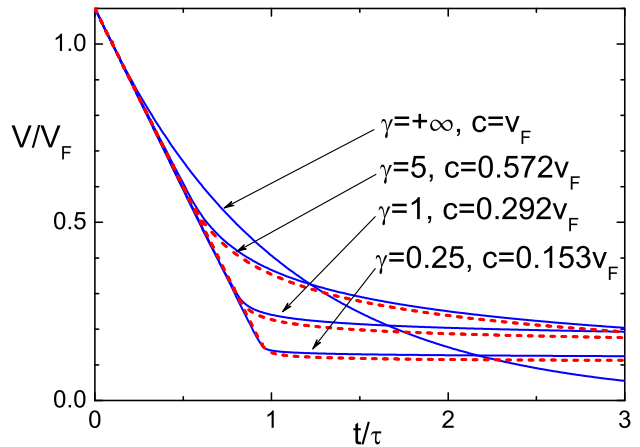}} 
\caption{\label{fig:Iv} (Color online) Decay of the ring current velocity of 1D bosons 
from the initial velocity of $1.1 v_{\rm F}$ at $t=0$. The solid (blue) and dashed (red) 
lines represent the results obtained with the approximate formula and ABACUS, 
respectively. The time scale is ${\tau}={N \pi \hbar^3}/{(2m g_{\mathrm{i}}^{2})}$. }
\end{figure}

\subsection{Currents in a ring}

In the presence of an obstacle a ring current can decay 
into supercurrent states with smaller momentum. Starting in one of the local minima of 
the spectrum in Fig.~\ref{fig:schem}, the kinetic energy of the center of mass will be 
transformed into elementary excitations above a lower supercurrent state,  conserving 
the total energy. The elementary excitations are quasiparticle-quasihole excitations in 
the Bethe-ansatz wave function \cite{lieb63:2} and may have mixed phonon and soliton 
character. Assuming that these have little effect on successive transitions, we estimate 
the decay of the center-of-mass velocity $v$ by the classical equation $Nm \dot{v} = 
-F_\mathrm{v}(v)$, where $F_\mathrm{v}$ is given by Eqs.~(\ref{dragf1D}) and 
(\ref{sq_our}). This  was integrated numerically and the result is shown in 
Fig.~\ref{fig:Iv}. At the initial supersonic velocity, where the drag force is saturated 
(see Fig.~\ref{fig:drag}) the supercurrent experiences constant deceleration. For 
$v\lesssim c$ the drag force decreases and the deceleration slows down. For the TG gas 
we find an analytical solution for exponential decay  $v(t)=v_0\exp(-t/\tau)$  for  
$v_0\leqslant v_\mathrm{F}$. In the weakly-interacting regime, the decay  becomes slow 
compared to experimental time scales.

\section{Conclusion}

Concluding, although the 1D Bose gas with finite repulsive inter-particle 
interaction shows superfluid phenomena of the {\em equilibrium type}, we show that in 
general its ability to support {\em dynamic} superfluid phenomena such as persistent ring 
currents is limited to a regime of very weak interactions; for a periodic potential, 
braking the gas, the persistent currents can be observed even in the TG regime at 
specific values of the velocity and density.

\acknowledgements
The authors thank L. Pitaevskii, A.J. Leggett and H.-P. B\"uchler for encouraging 
discussions and acknowledge hospitality of the Institut Henri Poincar\'e (Paris) and the 
Institute for Theoretical Physics of the University of Amsterdam. J.B. is supported by the Marsden 
Fund (Contract No. MAU0706) administered by the Royal Society of New Zealand.  J.S.C. and A.Y.C. 
acknowledge support from the FOM foundation of The Netherlands.

\end{document}